\documentstyle[pra,aps]{revtex}
\begin{document}
\draft
\title{Excited states of a dilute Bose-Einstein condensate in a harmonic trap}
\author{Alexander L. Fetter}
\address{Departments of Physics and Applied Physics, Stanford University,
Stanford, CA 94305-4060}
\author{Daniel Rokhsar}
\address{Department of Physics, University of California, Berkeley, CA
 94720-7300}
\date{\today}
\maketitle

\begin{abstract}
 The low-lying hydrodynamic normal modes of a dilute Bose-Einstein gas in
an isotropic
harmonic trap determine the corresponding Bogoliubov amplitudes.
In the Thomas-Fermi limit, these modes have large low-temperature
occupation numbers, and they permit an explicit construction  of the dynamic
structure function $S({\bf q},\omega)$.  The total noncondensate number
$N'(0)$ at zero temperature increases  like $R^6$, where $R$ is the
condensate radius measured in units of the oscillator length.
 The lowest dipole modes are constructed explicitly  in the Bogoliubov
approximation.
\end{abstract}


\section{Introduction}

Recent low-temperature experiments have observed Bose-Einstein
condensation in alkali atoms confined to harmonic
potentials\cite{And,Dav,hulet}. Subsequent investigations have detected
the first few low-lying collective modes of the  condensate~\cite{Jin,MOM}.
These results have stimulated a great deal of theoretical activity
pertaining to trapped dilute Bose systems.
Most of this work relies on the Bogoliubov approximation\cite{Bog,Yang},
which assumes that only a small fraction of the particles are excited
out of the condensate.  Clearly such a description fails completely
at the onset temperature $T_0$ for Bose condensation, where the
condensate occupation number $N_0(T)$ vanishes, but it does provide a very
useful description of the low-temperature behavior of a dilute condensed
Bose gas.

In a typical experiment of Refs.~\onlinecite{And,Dav}, the condensate
is ``large,'' in the sense that its diameter $R_0$ far exceeds the
characteristic width $d_0=\sqrt{\hbar/m\omega_0}$ of the single-particle
ground
state of the harmonic trap.   (Typically, $d_0 \sim 1\ {\rm \mu m}$ for the
traps of Refs.~\onlinecite{And,Dav}.) This result can be easily seen by
comparing
the relative contributions of  the kinetic energy, the trap potential
energy, and
the interaction potential energy  to the total energy of the gas.  The mean
kinetic energy
$\langle T\rangle_0 \sim N_0 (\hbar^2/2mR_0^2)$ diminishes as the
condensate grows, whereas both the mean single-particle confinement energy
$\langle V\rangle_0$ and the mean two-particle interaction energy
({\it i.e.,} the Hartree energy) $\langle V_H\rangle_0$ increase
with condensate size.  (The subscript $0$ denotes an expectation value
in the selfconsistent condensate.)  For a large condensate, the kinetic
energy is then much smaller than both the potential energies.
Neglecting the kinetic energy in comparison with the two potential
energies gives rise to the so-called ``Thomas-Fermi'' (TF) approximation,
which provides a simple description\cite{BP} of the spatially varying
condensate density $n_0(\bf r)$.

The two-body repulsion between atoms can be characterized by an $s$-wave
scattering length $a>0$, or equivalently an $s$-wave pseudopotential
$g = 4\pi \hbar^2 a/m$.  The characteristic Hartree energy of the
condensate is then $\langle V_H \rangle_0 \sim g N_0^2/R_0^3\sim
\hbar^2aN_0^2/mR_0^3$.  In the Thomas-Fermi limit, this Hartree energy must
dominate the kinetic energy $\langle T \rangle_0$.  Their ratio defines an
important  dimensionless quantity
\begin{equation}
\eta_0 \equiv \frac{N_0 a}{d_0} \sim
\frac{\langle V_H \rangle_0}{\langle T \rangle_0}.
\end{equation}
The TF limit therefore requires not only a macroscopically occupied
condensate ($N_0\gg 1$), but also the more stringent condition \cite{BP}
$\eta_0 = N_0a/d_0 \gg 1$.

In the Thomas-Fermi limit, the characteristic radius $R_0$ of the
condensate
is determined by the balance of $\langle V_H \rangle_0\sim
\hbar^2aN_0^2/mR_0^3$ and the trap
potential $\langle V \rangle_0$, which varies as
$\sim N_0 (m \omega_0^2 R_0^2)$ for a harmonic confining potential.
Minimizing $\langle V+V_H\rangle$ with respect to $R_0$ yields $N_0
\sim R_0^5
/(d_0^4a)$.  Introducing the dimensionless condensate radius
$R \equiv R_0/d_0$,
we find
$\eta_0 \sim R^5$ in  the Thomas-Fermi limit.

The collective spectrum of a confined cloud of Bose condensed atoms
differs qualitatively from that of a droplet of liquid helium of
comparable dimensions.  The helium droplet has nearly uniform density
except near a thin surface layer.  Its  bulk collective modes are
simply the compressional sound modes of a uniform fluid, whose quantization
is set by the boundary conditions at the surface of the drop.
For a drop of radius $R_0$, the minimum wavenumber of a phonon varies as
$1/R_0$, and the energy spacing between the collective modes is therefore
of order $\hbar s/R_0$, where $s$ is the bulk speed of sound.
As the drop size grows, the minimum excitation energy tends to zero
for a large drop, and the spectrum approaches the gapless spectrum
of bulk helium, as expected.

In a confined compressible atomic gas, however, the density of the
condensate varies with the size of the cloud, in contrast with
the nearly constant density of a drop of liquid.  We have seen that,
in the Thomas-Fermi limit, a cloud of radius $R_0$ contains
$N_0 \sim R^5(d_0/a)$ condensed atoms.  The mean density of the
condensate is then $N_0/R_0^3 \sim R^2$.

The speed of sound in a dilute Bose gas varies as the square root of its
density\cite{Bog}, so that $s \sim R$.  The typical energy spacing
between quantized sound modes in a harmonically confined cloud then varies
as $s/R \sim R^0$.  Thus while the maximum wavelength grows linearly
with the size of the cloud, the increased density of a larger cloud
raises the speed of sound proportionally, and the minimum excitation
frequency is independent of the radius of the cloud\cite{BP,Str}.

The excited states of spatially inhomogeneous Bose condensates can
be studied by two equivalent approaches.
The original work of Bogoliubov\cite{Bog} emphasized the underlying
quantum character of the problem.   His treatment of the uniform condensate
is readily generalized to the nonuniform case to yield the Bogoliubov
equations \cite{AP,deG} that constitute  the quantum-mechanical
Schr\"odinger equations for a pair of coupled amplitude functions.
The eigenvalues of this problem are the energies of the elementary
excitations of the Bose condensate; the corresponding amplitudes determine
the spatially varying noncondensate density.

More recently, a hydrodynamic approach has proved valuable in determining
both the excitation frequencies and the normal-mode amplitudes for the
low-lying excited states of a Bose condensate in a harmonic trap in the
Thomas-Fermi limit \cite{Str}.
It is not difficult to prove that these two descriptions are completely
equivalent in the Bogoliubov approximation\cite{ALF,WG}.
The present work exploits this feature to determine the contribution of
the low-lying collective modes to the zero-temperature noncondensate
density. In particular, the occupation number of these low-lying modes
is large in the TF limit, and a reasonable cutoff for the sum over all
such modes suggests that the total noncondensate number $N'$ then scales
as $R^6$ in this limit.  Since $N_0$ scales with $R^5$, the Bogoliubov
approximation that $N'\ll N_0$ necessarily fails for sufficiently large
$R$ (or $N$).

We review the basic formalism in Sec.\ II and summarize the equivalence
between the Bogoliubov and hydrodynamic descriptions in Sec.\ III,
along with the physical properties of the noncondensate.
The TF solution for the hydrodynamic normal modes is reviewed briefly in
Sec.\ IV.  In Sec.\ V we use the resulting eigenmodes to
determine the corresponding Bogoliubov amplitudes and low-temperature
noncondensate occupation.  Finally, in Sec.\ VI, we use the eigenmodes to
construct the dynamic structure function $S({\bf q},\omega)$ in the TF
limit. An
Appendix contains an explicit construction  of the exact lowest dipole
modes of
an interacting Bose gas as well as the corresponding modes
in the Bogoliubov
approximation.

\section{Basic formalism}
We briefly review the Bogoliubov approximation for a nonuniform condensed
Bose gas, which was introduced independently by Gross \cite{EPG} and
Pitaevskii \cite{LPP} to study vortices and their excitations.  (In this
context, the condensate has a uniform density $n_0$ except in the immediate
vicinity of the vortex core.)

At low densities, the two-particle interaction potential may be replaced
by a short-range pseudopotential with
$V({\bf r}) \approx g\delta({\bf r})$,
where $g$ is expressed in terms of the $s$-wave scattering length
$a$ through
the relation $g = 4\pi a \hbar^2/m$ \cite{FW}.  The present work considers
only ``repulsive'' interactions with $g, a > 0$.

For a uniform condensate with density $n_0$, a small deformation of the
condensate wave function with spatial scale $\lambda$ involves a squared
gradient ({\it i.e.,} kinetic) energy $\sim\hbar^2/2m\lambda^2$.
It is useful to define a ``coherence'' (or ``correlation'') length
\begin{equation}
\xi \equiv \frac{1}{\sqrt{8\pi a n_0}}
\end{equation}
through the balance between this kinetic energy and the repulsive
interparticle potential energy $g n_0$.
The coherence length becomes arbitrarily large for an ideal Bose gas
({\it i.e.,} when $a \to 0$).

In the Bogoliubov approximation, the fractional depletion of the
condensate $N'/N$ is of order\cite{Bog,Yang} $\sqrt{n_0a^3}$.
For the Bogoliubov approximation to be valid, $N'/N$ must be small,
which implies the number of particles per scattering volume $n_0 a^3$
is much smaller than unity.  (This condition is strongly violated in
liquid ${}^4$He.)  Equivalently, we require that
$\xi/a \sim (n_0 a^3)^{-1/2}$ be much greater than unity.

The presence of an external trap introduces another length $R_0$
that characterizes the spatial size of the condensate.
If $\xi \gg R_0$, then the system resembles an ideal Bose gas with
negligible interactions.  If $\xi\ll R_0$, however, the system differs
qualitatively  from an ideal Bose gas.
In the experiments on Bose-condensed sodium atoms in Ref.~\cite{Mewes},
the scattering length is $a \sim 4.9~\rm nm$, the trap has a characteristic
oscillator length $d_0\sim 1.9\ \mu\rm m$, and there are
$N_0 \sim 5\times 10^6$ condensed atoms.
The equivalent mean (isotropic)  condensate radius $R_0 \sim 20\ \mu\rm m$
implies a central condensate number density
$n_0\sim 4\times 10^{20}~\rm m^{-3}$.
Then  $\xi \sim 0.14~{\rm \mu m} \ll R_0$, and the system is indeed a
dilute interacting Bose gas with $a\ll\xi\ll R_0$, definitely
far from ideal.

For simplicity, we consider here only a spherical harmonic trap, with
\begin{equation}
V({\bf r}) = \case1/2 m\omega_0^2r^2,
\end{equation}
and a characteristic length scale
\begin{equation}
d_0 = \sqrt{\frac{\hbar}{m\omega_0}}
\label{eq:osc}
\end{equation}
corresponding to the Gaussian width of the single-particle ground state
of a single particle of mass $m$ in the trap.
When the trap contains a large number $N_0$ of condensed particles,
their mutual repulsion causes the cloud of atoms to expand.  The actual
condensate density $\sim N_0/R_0^3$ is then much smaller than the simple
estimate $\sim N_0/d_0^3$.  Specifically, when the Thomas-Fermi parameter
$\eta_0\equiv N_0a/d_0$ is sufficiently large, the dimensionless radial
expansion factor $R\equiv R_0/d_0$ is\cite{BP} $(15\eta_0)^{1/5}$.
This reduction in the particle density (by a factor of order
$\eta_0^{-3/5}$) means that the system remains dilute for
$\eta_0^{2/5}(a/d_0)^2\ll 1$.  Put another way, $N_0$ must be much less
than $(d_0/a)^6$, which is $\sim10^{15}$ for Bose-condensed sodium atoms
in Ref.~\cite{Mewes}.

 A spatially nonuniform  Bose condensate is characterized by a condensate
wave function $\Psi({\bf r})$ that can be normalized to the total number
of condensate particles, {\it i.e.,} $\int d^3r\,|\Psi|^2 = N_0$.
Then $n_0({\bf r})  = |\Psi({\bf r})|^2$ is the condensate particle
density.  For a dilute Bose gas at low temperature, $\Psi$ obeys a
selfconsistent nonlinear Schr\"odinger equation known as the
Gross-Pitaevskii (GP) equation
\cite{EPG,LPP}
\begin{equation}
(T + V +V_H -\mu )\Psi = 0,
\label{eq:GP}
\end{equation}
where $T= -\hbar^2\nabla^2/2m$ is the kinetic-energy operator,
$V({\bf r})$ is the trap potential energy operator,
$V_H({\bf r}) = g|\Psi({\bf r})|^2 = g n_0({\bf r})$  specifies the mean
(Hartree) pseudopotential
due to the condensate, and $\mu$ is the chemical potential.
For a stationary condensate, $\Psi$ can be taken as real, but the
generalization to a complex condensate wave function
$\Psi = e^{iS}\,|\Psi|$ with superfluid velocity
${\bf v}_s = (\hbar/m)\,\nabla S$ is not difficult.
(Such a complex condensate could describe, for example, a
vortex\cite{AP,ALF,EPG,LPP}.)

If the mean condensate kinetic energy
$\langle T\rangle_0 \equiv  \int d^3r\,\Psi^*\,T\,\Psi$
is negligible compared to the potential energies
$\langle V_H\rangle_0$ and $\langle V\rangle_0$ \cite{BP},
{ which holds for $\eta_0 \gg 1$,}
then the Gross-Pitaevskii equation [Eq.~(\ref{eq:GP})] can be approximated
by its last three terms.  In this limit, for each spatial position $\bf r$,
either the condensate wave function vanishes or the condensate density
satisfies  the ``Thomas-Fermi'' approximation
\begin{equation}
V_H({\bf r}) = gn_0({\bf r}) =
\big[\mu-V({\bf r})\big]\,\theta\big[\mu-V({\bf r})\big],
\label{eq:TF}
\end{equation}
where $\theta(x) $ denotes the unit positive step function.  For a
spherical harmonic trap with oscillator length $d_0$ and oscillator
frequency \(\omega_0\), the TF condensate density is an inverted parabola
that vanishes beyond a dimensionless cutoff radius
$R \approx (15\eta_0)^{1/5}$ defined in terms of  the chemical potential
$\mu = \frac{1}{2} \hbar\omega_0 R^2$
\cite{BP}.

The validity of this TF approximation has been investigated both
numerically for various values of the dimensionless parameter $R$
\cite{EB,RHBE} and analytically\cite{DPS} through an expansion in
powers of the ``small'' parameter $1/R^4$.
This latter treatment shows that the TF approximation fails in a
thin surface region of thickness
$\sim d_0 (d_0/R_0)^{1/3}\sim d_0 R^{-1/3}$, where the formally negligible
correction terms eliminate the singularity of  $|d\Psi_{\rm TF}/dr|^2$
at the
condensate surface.  As a result, the condensate kinetic energy acquires a
logarithmic correction of order
$(\hbar\omega_0/2R^2)\ln R$.

In an ideal Bose gas at zero temperature, all the particles are condensed
in the single-particle ground state of the trap.
Repulsive interactions excite a (small) fraction $N'/N$ of the particles
out of the condensate, even at zero temperature.
These excited particles occupy the various normal-mode eigenstates that
satisfy the (linear) Bogoliubov equations\cite{Bog,AP,deG,ALF,LPP}
\begin{mathletters}
\label{Bog}
\begin{equation} {\cal L}u_j - V_H v_j= E_j u_j,
\label{eq:Boga}
\end{equation}
\begin{equation} -V_H u_j + {\cal L}v_j= -E_j v_j,
\label{eq:Bogb}
\end{equation}
\end{mathletters}
for the coupled eigenfunctions $u_j({\bf r})$ and $v_j({\bf r})$, and
the associated eigenvalues $E_j$.  Here,  $j$ denotes a complete set of
quantum numbers, and the operator ${\cal L}$ has the form
\begin{equation}
{\cal L} = T + V + 2V_H-\mu,
\label{eq:L}
\end{equation}
where we again restrict our attention to stationary condensates.
(The generalization to the case of nonzero superfluid velocity
is not difficult\cite{ALF}.)
In these Bogoliubov equations, the minus sign in the coupling terms is
conventionally chosen to ensure that the ratio of the two amplitudes is
positive for a uniform Bose gas, where plane waves are the appropriate
eigenfunctions: $u_ke^{i{\bf k\cdot r}}$ and $v_ke^{i{\bf k\cdot r}}$.
These coupled Schr\"odinger equations for the amplitudes $u_j$ and
$v_j$ are analogous to the multicomponent Dirac equation.
(The eigenvalues of the Bogoliubov equations come in $\pm$ pairs; the
eigenfunctions of these pairs are related by a simple symmetry relation
\cite{AP}.)

For a localized trapped condensate in an unbounded confining  potential,
the condensate density $n_0$ and  the two-particle Hartree potential
$V_H$ both vanish at infinity, and the amplitudes obey the usual
quantum-mechanical bound-state boundary condition that $u_j$ and $v_j$
vanish for $r \to\infty$.
Furthermore, for positive eigenvalues, the eigenfunctions can be chosen
to satisfy the orthonormality condition \cite{AP}
\begin{equation}
\int d^3r\,(u_j^*u_k-v_j^*v_k) = \delta_{jk}.
\label{eq:ortho}
\end{equation}

The (low-temperature) total noncondensate number $N'(T)$ is obtained
by summing over all eigenstates
\begin{equation}
N'(T) = {\sum_j}'\,N_j'(T),
\end{equation}
where the primed sum omits the lowest eigenstate (which simply describes
the condensate itself).
Here $N_j'(T)$ is the spatial integral of the corresponding
temperature-dependent noncondensate density
\begin{equation}
n_j'({\bf r}) \equiv |v_j({\bf r})|^2
+\big(|u_j({\bf r})|^2 + |v_j({\bf r})|^2\big)(e^{\beta E_j}-1)^{-1}.
\label{eq:noncon}
\end{equation}
The second term vanishes as $T \rightarrow 0$, leaving the integral of
first term as the mean occupation number at zero temperature:
\begin{equation}
N_j'(0) = \int d^3r\,|v_j({\bf r})|^2.
\end{equation}
At nonzero temperature, the condition $N_0(T) = N - N'(T)$
determines the total condensate number, but the present work emphasizes
the zero-temperature  limit.

It is helpful to recall briefly the special case of a uniform bulk
condensate with constant density $n_0$ \cite{Bog,FWtemp}, where the
confining potential $V$ is absent and the Hartree interaction energy
and condensate chemical potential are equal:
$V_H = \mu = g n_0$.  The eigenfunctions are plane waves
$\propto e^{i{\bf k\cdot r}}$, and the energy eigenvalues have the
familiar Bogoliubov form
\begin{equation}
E_k = \sqrt{2T_kV_H + T_k^2} \approx \cases{\hbar s k,& for
\(k\xi\ll 1\),\cr
\noalign{\smallskip}\hbar^2k^2/2m,& for \(k\xi \gg
1\),\cr}
\label{eq:Bog}
\end{equation}
where $T_k = \hbar^2k^2/2m$ is the kinetic energy and
$s =\sqrt{4\pi a\hbar^2n_0/m^2}=\hbar/\sqrt2 m\xi$
is the speed of compressional sound.

The corresponding ``coherence factors'' $u_k$ and $v_k$ determine the
mixing of the two components, and obey the bosonic normalization
condition $u_k^2 -v_k^2= 1$ for each $\bf k$.
At zero temperature, the noncondensate occupation of the ${\bf k}$th
plane-wave mode is simply
\begin{equation}
N_{k}'(0) =  v_k^2 = \frac{1}{2}\,\bigg(\frac{T_k+V_H}{E_k}
- 1\bigg) \approx
\cases{\sqrt{V_H/8T_k} \approx (2\sqrt2\,k\xi)^{-1},&for
\(k\xi \ll 1\),\cr\noalign{\medskip} V_H^2/4T_k^2\approx
\frac{1}{4}(k\xi)^{-4},&for \(k\xi\gg
1\).\cr}
\label{eq:nonc}
\end{equation}
The long-wavelength  singularity in $N_k^\prime(0)\propto k^{-1}$ is
integrable, and
the short-wavelength behavior ensures that the total noncondensate density
$n'=(2\pi)^{-3}\int d^3k\,v_k^2 = \frac{8}{3}n_0\sqrt{n_0a^3/\pi}$
is not only finite but  also small relative to the condensate density.
At low
temperature, the corresponding additional thermal occupation for the $\bf
k$th state has the form
$\Delta N_{k}'(T)\equiv N_k'(T)-N_k'(0)\propto T/k^2$.

\section{equivalence with hydrodynamic formalism}

We now return to the general inhomogeneous case.
To proceed, it is useful to note that the GP equation (\ref{eq:GP}) for
the condensate wave function can be rewritten with Eq.\ (\ref{eq:L})
as ${\cal L}\Psi = V_H\Psi$, which suggests the following transformation
of the Bogoliubov amplitudes
\begin{equation}
u_j = \frac{\Psi\,U_j}{\sqrt{N_0}}\qquad\hbox{and}\qquad v_j =
\frac{\Psi\,V_j}{\sqrt{N_0}}\label{eq:trans}.
\end{equation}
In particular, it is straightforward to verify that
\begin{equation}
{\cal L}u_j = \frac{\Psi}{\sqrt{N_0}}\,(\tilde T + V_H)U_j,
\end{equation}
where $\tilde T$ is a differential operator defined by
\begin{equation}
\tilde T f \equiv
-\frac{\hbar^2}{2m|\Psi|^2}\,\nabla\cdot\big(|\Psi|^2\nabla\,f\big)=
-\frac{\hbar^2}{2m\,n_0}\,\nabla\cdot\big(n_0\nabla\,f\big).
\end{equation}

To simplify (\ref{Bog}) further, define
\begin{equation}
F_j = U_j + V_j\qquad\hbox{and}\qquad G_j = U_j -V_j;
\end{equation}
Then the Bogoliubov equations can be rewritten exactly as
\begin{mathletters}
\label{FG}
\begin{equation}
\tilde TF_j =E_jG_j,\label{eq:FGa}
\end{equation}
\begin{equation}
(\tilde T + 2V_H)G_j = E_jF_j.\label{eq:FGb}
\end{equation}
\end{mathletters}
These two equations can be combined to give a {\it single} equation
for $G_j$:
\begin{equation}
(\tilde T^2 + 2\tilde T V_H)G_j =
E_j^2G_j.
\label{eq:G}
\end{equation}
The corresponding $F_j$ follows from Eq.\ (\ref{eq:FGb}).
Comparison with Eq.\ (\ref{eq:ortho}) shows that the normalization for the
$j$th eigenstate is simply
\begin{equation}
1 = \frac{1}{2N_0}\int d^3r\,|\Psi|^2\,(F^*_jG_j +F_jG^*_j)
= \frac{1}{N_0}\int
d^3r\,|\Psi|^2\,\Re(F^*_jG_j).
\label{eq:normj}
\end{equation}
The zero-temperature occupation of the $j$th excited state is then
[compare Eq.\ (\ref{eq:noncon})]
\begin{equation}
N_j'(0) = \frac{1}{4N_0}
\int d^3r\,|\Psi|^2\,|F_j - G_j|^2.
\label{eq:nonconj}
\end{equation}

To make contact with the hydrodynamic description of these same normal
modes, recall that the second-quantized operators for the fluctuations
in the density $\hat\rho'$ and velocity potential $\hat\Phi'$ are simply
linear combinations of the field operators $\hat\phi$ and
$\hat\phi^\dagger$ \cite{ALF}.  It follows that the corresponding
normal-mode amplitudes $\rho_j$ and $\Phi_j$ are also linearly related
to the Bogoliubov amplitudes $u_j$ and $v_j$.
A straightforward comparison yields
\begin{mathletters}\label{linab}
\begin{equation}
\rho_j = \frac{n_0}{\sqrt{N_0}}\,G_j,\label{eq:rhojj}\end{equation}
\begin{equation}\Phi_j = \frac{\hbar}{2im\sqrt{N_0}}\,F_j.
\end{equation}
\end{mathletters}
Conversely, if the hydrodynamic amplitudes are known, the corresponding
Bogoliubov wave functions
become
\begin{mathletters}
\label{linv}
\begin{equation}
u_j = \case 1/2 \Psi\bigg(\frac{\rho_j}{n_0} +
\frac{2im\,\Phi_j}{\hbar}\bigg) = \frac{\rho_j}{2\Psi} +
\frac{im\,\Psi\,\Phi_j}{\hbar},\label{linva}\end{equation}
\begin{equation} v_j = \case 1/2 \Psi\bigg(\frac{\rho_j}{n_0} -
\frac{2im\,\Phi_j}{\hbar}\bigg)= \frac{\rho_j}{2\Psi} -
\frac{im\,\Psi\,\Phi_j}{\hbar}.\label{linvb}\end{equation}
\end{mathletters}
A straightforward combination of Eqs.\ (\ref{FG}) and (\ref{linab})
immediately reproduces the known hydrodynamic equations for the normal
modes of a stationary condensate \cite{Str,ALF,WG}.  For example,
the product $V_HG_j$ is just the density-fluctuation normal-mode
eigenfunction $\rho_j$ itself (apart from a constant factor) and
Eq.\ (\ref{eq:G}) becomes
\begin{equation}-\frac{4\pi a\hbar^2}{m^2}\nabla\cdot
\big(n_0\nabla\rho_j\big)+
\frac{\hbar^2}{4m^2}\nabla\cdot
\bigg\{n_0\nabla\bigg[\frac{1}{n_0}\nabla\cdot
\bigg(n_0\nabla\,\frac{\rho_j}{n_0}\bigg)\bigg]\bigg\}=
\omega_j^2\rho_j,\label{eq:rhoj}\end{equation}
where $\omega_j = E_j/\hbar$ is the normal-mode eigenfrequency.

\section{low-lying normal-mode hydrodynamic amplitudes in TF limit}

It is convenient to rewrite the Bogoliubov equations in terms of
suitably rescaled variables.  We let ${\bf x} \equiv {\bf r}/R_0$ be
the dimensionless position vector, where $R_0$ is the characteristic
condensate radius, and introduce the dimensionless condensate wave function
\begin{equation}
\chi = \bigg(4\pi \tilde{\eta}_0 \frac{R_0^3}{N_0}\bigg)^{1/2} \Psi,
\label{chi-def}
\end{equation}
where the quantity in parentheses is roughly the volume per condensate
particle.  The rescaled condensate wave function $\chi$ satisfies the
dimensionless
radial normalization condition
$\int_0^{\infty} x^2\,dx\,|\chi|^2 = \tilde\eta_0$, where the
dimensionless parameter $\tilde{\eta}_0$ in Eq.~(\ref{chi-def})
is defined by
\begin{equation}
\tilde{\eta}_0 \equiv \frac{\eta_0}{R^5} = \frac{N_0 a d_0^4}{R_0^5}.
\end{equation}
In the Thomas-Fermi limit, $\tilde{\eta}_0$ becomes independent of the
size of the condensate.
To complete our scaling of variables, we express all energies in units
of $\hbar\omega_0$.

The three relevant operators then become
\begin{mathletters}
\label{dimen}
\begin{equation}
\tilde T = -\frac{1}{2R^2\,|\chi|^2}\,\nabla_{\bf x}
\cdot\big(|\chi|^2\nabla_{\bf x}\big),
\label{eq:dimena}
\end{equation}
\begin{equation}
V = \case 1/2 R^2\,x^2,
\label{eq:dimenb}
\end{equation}
\begin{equation}
V_H = R^2|\chi|^2.
\label{eq:dimenc}
\end{equation}
\end{mathletters}
These expressions clearly exhibit the dependence on the large parameter
$R^2$ and show that the trap and Hartree energies are comparable to
one another and  both much larger than the kinetic energy.
In the rescaled variables, the product $\tilde TV_H$ is independent of $R$,
and the basic eigenvalue Eq.\ (\ref{eq:rhoj}) for the hydrodynamic
normal-mode amplitude $\rho_j$ becomes
\begin{equation}
-\nabla_{\bf x}\cdot\big(|\chi|^2\nabla_{\bf x}\,\rho_j\big) +
\frac{1}{4}\,\epsilon\,\nabla_{\bf x}\cdot\bigg\{|\chi|^2\nabla_{\bf
x}\bigg[\frac{1}{|\chi|^2}\nabla_{\bf x}\cdot\bigg(|\chi|^2\nabla_{\bf
x}\,\frac{\rho_j}{|\chi|^2}\bigg)\bigg]\bigg\} = E_j^2\rho_j,
\label{eq:Strin}
\end{equation}
where $E_j$ is the dimensionless energy (or frequency) of the $j$th normal
mode, and $\epsilon\equiv R^{-4}$ is the appropriate small expansion
parameter.

For a spherical trap, Baym and Pethick \cite{BP} showed that
 $\tilde{\eta}_0 \approx \frac{1}{15}$ in the
TF limit, and the rescaled condensate wave function
is simply\cite{BP}
\begin{equation}
|\chi|^2 \approx \chi_0^2 = \frac{1}{2}(1-x^2)\,\theta(1-x),
\label{eq:chi0}
\end{equation}
where $x$ is the scaled radial variable.
For a spherical trap, the eigenfunctions of Eq. (\ref{eq:Strin})
can be written
as a product of (real) radial functions $\rho_{nl}(x)$ and a
spherical harmonic $Y_{lm}(\theta,\phi)$, where $j = (nlm)$,
and $n$ is the
radial quantum number.

Stringari \cite{Str} has solved Eq.\ (\ref{eq:Strin}) in the TF limit
({\it i.e.,} to zeroth order in $\epsilon$) and has shown that the
eigenfunctions
and eigenvalues are independent of $R$, as discussed in Sec.~I.
In particular,
the TF energy eigenvalue is given by
\begin{equation}
E_{nl}^2 = l+n(2n + 2l + 3) = \alpha - \case 1/2 +2n(n+\alpha + 1).
\label{eq:eigvl}
\end{equation}
where $\alpha\equiv l+\frac{1}{2}$ is half an odd integer.
The corresponding  radial  functions
$\rho_{nl}(x)$ have the form
$x^lP_{nl}(x^2)$, where
$ P_{nl}(x^2)$  are
$n$th-order polynomials
in
$x^2$.  It is not difficult to verify that these polynomials satisfy the
hypergeometric equation \cite{AS} with the explicit (unnormalized) form
\begin{equation}
P_{nl}(u) = F(-n,n+\alpha +1;\alpha+1;u) =
\frac{\Gamma(\alpha+1)}{\Gamma(n+\alpha+1)}\,u^{-\alpha}\,
\frac{d^n}{du^n}\,
\big[\,u^{n+\alpha}\,(1-u)^n\,\big];
\label{eq:poly}
\end{equation}
the first two such polynomials are $P_{0l}=1$ and $P_{1l} =
1-(\alpha+2)u/(\alpha+1) = 1 - (2l+5)u/(2l+3)$.

The polynomials $P_{nl}$ are a special class of Jacobi polynomials, with
$P_{nl}(u) \propto P_n^{(\alpha,0)}(1-2u)$.  For any $l$, they form an
orthonormal set on the interval $0\le u\le 1$ with weight $u^\alpha$.
An $n$-fold integration by parts with the explicit differential expression
in Eq.\ (\ref{eq:poly}) readily yields the radial normalization integral
\begin{equation}
I_{nl}^0 = \int_0^1
x^2\,dx\,\big[\rho_{nl}(x)\big]^2 =\case 1/2
\int_0^1 du\,u^\alpha\,\big[P_{nl}(u)\big]^2 =
\bigg[\frac{\Gamma(\alpha+1)\,n!}{\Gamma(n+\alpha+1)}\bigg]^2\,
\frac{1}{2(2n +\alpha + 1)}\,,
\label{eq:norm}
\end{equation}
which also follows directly from the properties of the Jacobi polynomials
\cite{AS}.

The transformation introduced in Eq.~(\ref{eq:trans}) explicitly eliminates the
chemical potential from the Bogoliubov equations for the amplitudes
$u_j$ and
$v_j$.  As shown above, it leads to the hydrodynamic Eq.~(\ref{eq:Strin})
that
involves only the condensate density.  In principle, the Bogoliubov
description is
wholly equivalent to this hydrodynamic description, but an explicit
verification involves
higher-order terms in the condensate wave function\cite{FF}.

To understand the situation in more detail,
recall that, in the grand canonical ensemble  at zero temperature, the
chemical potential $\mu$
determines both the total number of particles $N$ and the number $N_0\le N$
in the
condensate\cite{FW-bose}. In the present approximation
 of retaining
only the condensate contribution ({\it i. e.\/} $N\approx N_0$), the
Gross-Pitaevskii equation
wholly characterizes the resulting
 functional
dependence $N_0(\mu)$.  For large dimensionless $\mu$, this relation is
just the
familiar TF result
plus small corrections\cite{FF}.
It is straightforward  to verify that  direct substitution  of the
TF chemical potential and the TF wave function $\chi_0$ from
Eq.~(\ref{eq:chi0})   into the
Bogoliubov equations  yields an incorrect
eigenvalue spectrum. A more careful treatment that includes  the leading
correction
$\chi_1$   of order
$1/R^4$
indeed reproduces the hydrodynamic eigenvalues and eigenfunctions.

\section{noncondensate occupation at low temperature}

Stringari's theoretical prediction \cite{Str} of the frequencies of
the lowest hydrodynamic normal modes rapidly received experimental
confirmation \cite{Jin,MOM}.
The present work extends his results to determine the occupation number
of the same low-lying normal modes.  This analysis makes essential use of
the quantum-mechanical Bogoliubov amplitudes $u_{nlm}$ and $v_{nlm}$
for quasiparticle creation and annihilation operators associated with
these particular single-particle states \cite{ALF,WG}, and exploits
the equivalence between the hydrodynamic and Bogoliubov descriptions.

Equation (\ref{linab}) expresses the radial amplitude $G_{nl}$ directly in
terms of
$\rho_{nl}$;  apart from a constant factor, we have $G_{nl}(x)\propto
\rho_{nl}(x)/|\chi_0(x)|^2$, where $|\chi_0(x)|^2= \frac{1}{2}(1-x^2)$ is
the parabolic TF condensate
density profile.  In addition, Eqs.~(\ref{eq:FGb}) and (\ref{eq:dimenc})
show that $F_{nl} \approx
2R^2|\chi_0|^2G_{nl}/E_{nl}$, apart from corrections that become small as
$R\to \infty$.  Together, these
expressions suggest the following normalization
\begin{mathletters}
\label{normBog}
\begin{equation}
F_{nl}(x) = \frac{2R\,C_{nl}}{E_{nl}}\,\rho_{nl}(x)
=\frac{2R\,C_{nl}}{E_{nl}}\,x^l\,P_{nl}(x^2),
\label{normBoga}
\end{equation}
\begin{equation}
G_{nl}(x) = \frac{C_{nl}}{R}\,\frac{\rho_{nl}(x)}{|\chi_0(x)|^2}
=\frac{C_{nl}}{R}\,\frac{x^l\,P_{nl}(x^2)}{|\chi_0(x)|^2},
\label{eq:normBogb}
\end{equation}
\end{mathletters}
where $C_{nl}$ is a normalization constant determined from
Eq.\ (\ref{eq:normj}).  A combination of the previous results leads to
the explicit expression
\begin{equation}
C_{nl}^2 = \frac{2\pi\tilde\eta_0\,E_{nl}}{I_{nl}^0},
\label{eq:C}
\end{equation}
which completely determines the radial amplitudes $F_{nl}$ and $G_{nl}$
in the TF limit.  Note that while $F_{nl}$ is of order $R$, $G_{nl}$ is
of order $1/R$, so that $U_{nl} = \case 1/2 (F_{nl} + G_{nl})$
and $V_{nl} = \case 1/2 (F_{nl} - G_{nl})$ are {\it both} large for
the low-lying states of a large condensate.
[A similar behavior occurs at long wavelengths ($k\xi\lesssim 1$) for a
uniform dilute Bose gas, as seen from the bosonic  normalization
condition $u_k^2-v_k^2 = 1$  and Eq.  (\ref{eq:nonc}).]

The zero-temperature occupation of the low-lying normal modes with quantum
numbers $nlm$ follows directly from Eq.\ (\ref{eq:nonconj}).  The leading
term from $|F_{nl}|^2$ is of order  $R^2$, and the cross term between
$F_{nl}$ and $G_{nl}$ just reproduces the normalization integral from
Eq.\ (\ref{eq:norm}).  It is convenient to introduce the general class of
integrals
\begin{equation}
I_{nl}^j \equiv \int_0^1 x^2\,dx\,
(1-x^2)^j\,\big[\rho_{nl}(x)\big]^2=\case 1/2 \int_0^1
du\,u^\alpha\,(1-u)^j\,\big[P_{nl}(u)\big]^2.
\label{eq:Ij}
\end{equation}
For $j=0$ this quantity is simply the normalization integral considered
previously; for $j=1$, it can be evaluated with repeated integration by
parts, leading to the ratio
\begin{equation}
\frac{I_{nl}^1}{I_{nl}^0} = \frac{2n(n+\alpha+1) +
\alpha}{(2n+\alpha)(2n + \alpha + 2)}= \frac{E_{nl}^2 + \case
1/2}{(2n+\alpha)(2n + \alpha + 2)}.
\label{eq:I1}
\end{equation}
A combination of these results yields an explicit expression for the
zero-temperature occupation of the low-lying excited states of a large
isotropic condensate in an isotropic harmonic trap:
\begin{equation}
N_{nl}'(0)\approx \frac{R^2}{4E_{nl}}\,\frac{E_{nl}^2 + \case
1/2}{(2n+\alpha)(2n + \alpha + 2)} -  \case 1/2,
\label{eq:N'}
\end{equation}
where the omitted term comes from the radial integral of $|G_{nl}|^2$
and is of order $R^{-2}$.

Since $R^2\gg 1$ in the TF limit, the noncondensate occupation $N_{nl}'(0)$
is large for the low-lying states with small $n$ and $l$, whose
dimensionless energy is
of order unity.  For example, for the lowest dipole mode with $n=0$, $l=1$,
we have $E_{01} = 1$ and $N_{01}'(0)\approx \frac{1}{14}R^2 - \case 1/2$.
This behavior is evidently very similar to that for a uniform dilute Bose
gas, where Eq.\ (\ref{eq:nonc}) shows that
$N_k'(0)= (2\sqrt2\,k\xi)^{-1} \gg 1$
for long wavelengths such that $k\xi \lesssim 1$.
Indeed, Eq.\ (\ref{eq:nonc}) also makes clear that the long-wavelength
approximation fails when the kinetic energy $T_k$ of the plane-wave
state becomes comparable to the Hartree energy $V_H$, and a similar
behavior is expected in the present case of a trapped condensate.

Specifically, the amplitude $F_{nl}$ was obtained from $G_{nl}$ by
neglecting $\tilde T$ relative to $2V_H$ in Eq.\ (\ref{eq:FGb}).
This approximation holds for radial states with sufficiently few nodes,
but it necessarily fails for highly excited states whose large kinetic
energy reflects the bending energy associated with rapid oscillations
(and hence many nodes) in the wave function \cite{Str}. By analogy
with the corresponding situation for a uniform Bose gas, it is natural
to conjecture that the explicit expression in Eq.\ (\ref{eq:N'}) is
valid only for low-lying modes with $N_{nl}'(0)\gtrsim 1$.
Verification of this conjecture would require a detailed study of the
highly excited modes of the large condensate;  it involves corrections
to the TF condensate wave function associated with the boundary layer
\cite{DPS} and a WKB (phase-integral) description of the rapid
oscillations inherent in the short-wavelength limit.
This difficult analysis remains for  future investigation.

The total number $N'$ of noncondensed particles is the sum of $N_j'$
over all eigenstates of the Bogoliubov equations, omitting the lowest
solution with zero energy (which describes the condensate itself).
In the present case of a spherical condensate at zero temperature, we have
\begin{equation}
N'(0) = {\sum_{nl}}'\,(2l+1)\,N_{nl}'(0),
\label{double-sum}
\end{equation}
where the factor $2l+1$ represents the degeneracy associated with the sum
over azimuthal quantum numbers $m$ and the prime on the sum indicates
that the term $n = l = 0$ is omitted.  In the TF limit, the
zero-temperature occupation of the low-lying modes is given by Eq.\
(\ref{eq:N'}).  As argued above, the sum must be cut off when $N_{nl}'(0)
\approx 1$.
(Apart from a numerical factor of order unity, an analogous cutoff
gives the correct total noncondensate fraction for a uniform condensate
at zero
temperature,
$N'/N_0 \sim \sqrt{n_0a^3}$).
Stringari \cite{Str} has suggested that the TF expression for $E_{nl}$
in Eq.~(\ref{eq:eigvl}) holds for $E_{nl}\lesssim \mu$, which is
$\frac{1}{2}R^2$ in the TF limit.
It is not hard to see that this criterion provides a qualitatively
similar cutoff.

For large $R$, the $N_{nl}'$ vary slowly with $n$ and $l$, and the
double sum Eq. (\ref{double-sum}) can be approximated by an integral
over continuous variables $n$ and $\alpha \equiv l + \case 1/2$, with
\begin{equation}
N'(0) =  {\sum_{nl}}'(2l+1)\,N_{nl}'(0)\approx
\int dn\int d\alpha\,2\alpha \,
\bigg[\frac{R^2}
{4E_{nl}}\,\frac{E_{nl}^2 + \case 1/2}{(2n+\alpha)
(2n + \alpha + 2)}- \case 1/2\bigg],
\end{equation}
where this double integral runs over the region $N_{nl}' \ge 1$.
To clarify its structure, it is convenient to introduce new variables
\begin{equation}
s \equiv n + l + 1 = n+\alpha+\case 1/2
\qquad\text{and}\qquad
t \equiv n+ \case 1/2,
\end{equation}
or, equivalently
\begin{equation}
n = t-\case 1/2
\qquad\hbox{and}\qquad
\alpha= s-t;
\end{equation}
this transformation has unit Jacobian, and the allowed region in the $st$
plane is the first octant $0 \le t\le s$, apart from a small region
around the origin.  In these new variables, the TF energy eigenvalue in
Eq.\ (\ref{eq:eigvl}) becomes $E^2 = 2st - 1$, and the
corresponding noncondensate occupation number is
\begin{equation}
N_{nl}'(0)\approx \frac{R^2}{4}\,\frac{2st-\case 1/2}{\sqrt{2st
-1}}\,\frac{1}{(s+t)^2-1}-\case 1/2.
\label{eq:N''}
\end{equation}
To isolate the dominant contribution to the integral in the  $st$ plane,
it is helpful to introduce plane-polar coordinates ($\zeta,\phi$),
with $s= \zeta\cos\phi$ and $t=\zeta\sin\phi$;
for large $\zeta$, the leading behavior is $N'\sim R^2/\zeta$, apart from
angular factors.  It is not difficult to see that the angular integral over
$0\le \phi\le\pi/4$ converges, and the radial integral must be cut off at
$\zeta_{\rm max} \sim R^2$.
In this way, the total noncondensate number at zero temperature becomes
\begin{equation}
N'(0)\sim\int_1^{R^2}
\zeta^2\,d\zeta\,\frac{R^2}{\zeta}\sim R^6.
\label{eq:N'sim}
\end{equation}

As shown by Baym and Pethick \cite{BP}, the number of condensed particles
in the TF limit is given by $N_0 \approx \frac{1}{15} d_0R^5/a$,
proportional to $R^5$.  So from Eq. (\ref{eq:N'sim}), the ratio of
uncondensed to condensed particles at zero temperature in the Thomas-Fermi
limit is
\begin{equation}
\frac{N'(0)}{N_0} \sim R \frac{a}{d_0},
\end{equation}
apart from a numerical constant of order unity.  As expected, this is
comparable to the uncondensed fraction in a homogeneous Bose gas of
density $N_0/R_0^3$.
Since the validity of the Bogoliubov approximation depends on
the condition $N'/N_0 \ll 1$, we find that the present description
holds only for $R \ll d_0/a$.
When combined with Baym and Pethick's result, this condition
indicates that the condensate is dilute only for
$N_0 \sim R^5 (d_0/a) \ll (d_0/a)^6$, in agreement with the condition
found in Sec.\ II.

It is evident from Eqs.~(\ref{eq:ortho}) and (\ref{eq:noncon}) that the
zero-temperature eigenvalues and eigenfunctions of the Bogoliubov equations also
determine the  low-temperature thermal depletion of the condensate
through the
relation
\begin{equation}
\Delta N_{nl}'(T) = N_{nl}'(T) -N_{nl}'(0)\approx \frac {1 +
2N_{nl}'(0)}{\exp(\beta\hbar\omega_0 E_{nl})-1},\end{equation}
where $E_{nl}=\sqrt{l + n(2n+2l+3)}$ is the dimensionless energy
eigenvalue, here
taken from Eq.~(\ref{eq:eigvl}).  It is convenient to define a
characteristic
temperature $\Theta \equiv \hbar\omega_0/k_B$ for thermal excitation
of the
first
excited state; note that  $\Theta$
is much smaller than the ideal-gas transition temperature
$T_0^{(0)} \sim \Theta
N^{1/3}$.  Furthermore, the actual
transition temperature $T_0$ for a dilute trapped Bose gas
 is only slightly less than the ideal-gas transition temperature
$T_0^{(0)}$, so
that
$\Theta/T_0\approx N^{-1/3}$.  As a
result, the low-temperature thermal occupation of each low-lying mode
increases
linearly with temperature
\begin{equation}\Delta N_{nl}'(T) \approx [1 +2N_{nl}'(0)]
\frac{T}{\Theta\,E_{nl}}
\end{equation}
for $\Theta \,E_{nl}\ll T\ll T_0$.

\section{dynamic structure factor $S(\lowercase {\bf q},\omega)$}

Consider an external probe that scatters with momentum transfer
$\hbar\bf q$ and energy transfer $\hbar\omega$ to a target.  If the
probe  couples weakly to the number density of the target (here,  the
trapped Bose
condensed system), the differential cross section $d^2\sigma/d\Omega
d\omega $ is
proportional to   the dynamic structure factor\cite{PN,AM}
\begin{equation}S({\bf q},\omega) =
\frac{1}{NZ}\,\sum_{fi}e^{-\beta E_i}\,|\langle f|\tilde\rho_{\bf
q}'^\dagger|i\rangle|^2\,\delta\bigg(\omega-\frac{E_f-E_i}{\hbar}\bigg),
\label{eq:dyn}\end{equation}
where $i$ and $f$ refer to exact states of the interacting
target with energies $E_i$ and $E_f$, $Z=\sum_i \exp(-\beta E_i)$ is
the target partition function, and
$\tilde\rho_{\bf q}'^\dagger = \int d^3r\,e^{i{\bf q\cdot
r}}\,\hat\rho'({\bf r})$ is the
``creation operator'' for a density fluctuation with wave number $\bf
q$.

In the Bogoliubov approximation, the density-fluctuation operator
$\hat\rho'({\bf
r})$ is proportional to a linear combination of the field operators
$\hat\phi({\bf
r})$ and $\hat\phi^\dagger({\bf r})$.  As a result,  Eq.~(\ref{eq:rhojj})
immediately yields the corresponding expansion in bosonic quasiparticle
operators\cite{ALF}
$\alpha_j$ and $\alpha_j^\dagger$
\begin{equation}\hat\rho'({\bf r}) ={\sum_j}'\big[\rho_j({\bf r})\alpha_j
+ \rho_j^*({\bf r})\alpha_j^\dagger\big],\end{equation}
where $\rho_j$ is essentially a linear combination of the Bogoliubov
amplitudes,
and the primed sum runs over all the excited states of the condensate.
The
evaluation of the dynamic structure factor is straightforward, giving the
explicit
result
\begin{equation} S({\bf
q},\omega)=\frac{1}{N}\,{\sum_j}'\,\big[(1+f_j)\,|\tilde\rho_j^*({\bf
q)}|^2\delta(\omega-E_j/\hbar)+f_j\,|\tilde\rho_j({\bf
q)}|^2\delta(\omega+E_j/\hbar)\big],\label{eq:S}\end{equation}
where $f_j = \big[\exp(\beta E_j) -1\big]^{-1}$ is the thermal
Bose-Einstein
function and $\tilde\rho_j({\bf q})$ is the spatial Fourier transform of
$\rho_j({\bf r})$.

For a spherical trap, the amplitudes are given in Eq.~(\ref{eq:normBogb}).
Introducing the dimensionless variable $Q =qR_0$, we readily obtain the
following
dimensionless dynamic structure factor (scaled with the oscillator
frequency $\omega_0$)
\begin{equation}S(Q,\omega)\approx
\frac{1}{2\tilde\eta_0R^2}\,{\sum_{nl}}'\,\frac{(2l+1)E_{nl}}{I_{nl}^0}\,
|p_{nl}(Q)|^2\,
\big[(1+f_{nl})\,\delta(\omega-E_{nl})
+ f_{nl}\,\delta(\omega+E_{nl})\big],
\label{eq:dyndim}\end{equation}
where $E_{nl}$ is the dimensionless energy (or frequency) from
Eq.~(\ref{eq:eigvl}) and
\begin{equation}p_{nl}(Q) = \int_0^1
dx\,x^{l+2}\,P_{nl}(x^2)\,j_l(Qx).\label{eq:pq}\end{equation}
 Equation (\ref{eq:dyndim}) makes it easy to verify that this approximate
dynamic structure factor obeys the
detailed-balance condition\cite{PN}
$S(-{\bf Q},-\omega) = e^{-\beta\hbar\omega}S({\bf Q},\omega)$.

The integral in Eq.~(\ref{eq:pq}) can be evaluated
 with the  standard expression for the spherical Bessel function $j_l$
as an
$l$-fold derivative of
$j_0$\cite{Bess}
\begin{equation}\frac{j_l(x)}{x^l} =
(-2)^l\,\frac{d^l}{du^l}j_0(x),\end{equation}
where $u = x^2$, along with the explicit formula for $P_{nl}(x^2)$ in
Eq.~(\ref{eq:poly}):
\begin{equation}p_{nl}(Q) =
\frac{\Gamma(\alpha+1)}{\Gamma(n+\alpha+1)}\,\frac{2^{l-1}(-1)^l}{Q^l}\,
\int_0^1du\,\frac{d^n}{du^n}\big[u^{n+\alpha}(1-u)^n\big]\,
\frac{d^l}{du^l}j_0(Qx).\label{eg:Ray}\end{equation}
An $n$-fold integration by parts then yields
\begin{equation}p_{nl}(Q) =
\frac{\Gamma(\alpha+1)}{\Gamma(n+\alpha+1)}\,\frac{Q^{2n+l}}{2^{n+1}}\,
\int_0^1du\,u^{n+\alpha}\,(1-u)^n\>\frac{j_{n+l}(Qx)}{(Qx)^{n+l}},
\label{eg:Ray1}\end{equation}
which can be integrated term-by-term after expanding the spherical Bessel
function as a power series in $uQ^2$.
The resulting series can be re-summed to yield the remarkably simple
expression
\begin{equation}p_{nl}(Q) =
\frac{\Gamma(\alpha+1)\,n!}{\Gamma(n+\alpha+1)}\,\frac{j_{2n+l+1}(Q)}{Q}.
\label{eq:pnl}\end{equation}

It is convenient to rewrite the dynamic structure factor as
\begin{equation}S(Q,\omega)={\sum_{nl}}'S_{nl}(Q)\,
\big[(1+f_{nl})\,\delta(\omega-E_{nl})
+ f_{nl}\,\delta(\omega+E_{nl})\big],\label{eq:dynstr}\end{equation}
 where $f_{nl} = [\exp(\Theta E_{nl}/T)-1]^{-1}$, and
\begin{equation}S_{nl}(Q) =
\frac{(2l+1)\,(2n+l+\frac{3}{2})\,E_{nl}}{\tilde\eta_0Q^2R^2}\,
\big[j_{2n+l+1}(Q)\big]^2.\label{eq:dynstr1}\end{equation}
For $Q\to 0$, the leading term arises from  the dipole-sloshing mode $n=0$,
$l=1$, with $S_{01} \approx Q^2/2R^2
+{\cal O}(Q^4)$;  the next contributions (of order $Q^4$) arise from the
terms with $n=1$, $l=0$ and $n=0$,
$l=2$.   In addition, the frequency integral is simply the
static structure factor
\begin{equation}\int_{-\infty}^{\infty}d\omega\,S(Q,\omega) = S(Q) =
{\sum_{nl}}'S_{nl}(Q)\,\coth\bigg(\frac{\Theta
E_{nl}}{2T}\bigg),\label{eq:Sq}\end{equation}
with the long-wavelength limit
\begin{equation}S(Q) \approx
\frac{Q^2}{2R^2}\,\coth\bigg(\frac{\Theta}{2T}\bigg) + {\cal
O}(Q^4).\end{equation}
Finally, the first moment is the $f$-sum rule\cite{PN}
\begin{equation}\int_{-\infty}^{\infty}d\omega\,\omega\,S(Q,\omega) =
{\sum_{nl}}'S_{nl}(Q)\,E_{nl} = \frac{Q^2}{2R^2}\>\>
\big(=\frac{\hbar^2q^2}{2m}
\>\>\hbox{in conventional units}\big).\label{eq:fsum}\end{equation}
The dipole-sloshing mode exhausts this sum rule at long wavelengths,
but it
otherwise implies a rather
intricate identity involving sums of squares of spherical Bessel functions;
it is easy to verify this relation
through terms of order
$Q^4$, but we have not sought an independent derivation.

\section{discussion}

This work has shown how the known hydrodynamic amplitudes\cite{Str} for a
dilute condensed Bose gas in a
spherical harmonic trap can provide the corresponding Bogoliubov spatial
amplitudes for the quantum-mechanical
field operators, including their absolute normalization.  These normalized
amplitudes in turn determine
several important  physical quantities, such as the excitation of each
normal
mode, both in the ground state and  at low temperatures ($T\ll T_0$), and
the dynamic structure factor that
describes the inelastic scattering of a weakly interacting probe that
couples to the density fluctuations of the
Bose condensate.  The hydrodynamic amplitudes are accurate only for
low-lying modes, for their derivation  neglects
the quantum-mechanical kinetic energy that becomes increasingly important
at short wavelengths.  Thus our
estimate of the total condensate depletion   involves  a cut-off  that
explicitly omits all the high-lying normal
modes.

As noted by Singh and Rokhsar\cite{SR} and Stringari\cite{Str}, the exact
eigenfrequencies of the lowest dipole mode $l = 1$ for an anisotropic
harmonic trap coincide with the bare oscillator frequencies;  numerical
work by several groups \cite{SR,Edw,EDCB,Rup,You} has confirmed this
conclusion for moderate values of $N_0$.  The Appendix contains an
analytical proof in the Bogoliubov approximation; it provides an
explicit construction of the lowest dipole states for any solution of
the GP equation, including those containing one or more vortices.

\acknowledgements
We are grateful to S.\ Bahcall, M.\ Cole, A.\ Griffin, and S.\ Stringari
for
valuable comments and suggestions. This work was supported in part
by the National Science Foundation, under Grants No.~DMR 94-21888
and DMR 91-57414.

\appendix
\section*{dipole mode}

Recently, the collective excitation spectrum of harmonically confined Bose
gases has been calculated by several groups \cite{Str,SR,Edw,EDCB,Rup,You}.
It has been noted \cite{Str,SR} that these spectra contain a trio of
exact collective modes that correspond to the simple-harmonic oscillation
of the center of mass of the condensate.  If the interacting condensate is
displaced without deformation, the interparticle interactions are unchanged.
Thus, the system experiences a restoring force that is simply linear in the
displacement, with spring constant equal to that of the bare trap.
There is one such mode for each of the three Cartesian directions.

In this Appendix, we construct the exact many-body raising operator that
creates these ``dipole'' or ``sloshing'' modes and demonstrate the
independence of their frequencies on the interparticle interactions.
Since the collective modes of the dilute Bose gas are expected to be
well-described by the Bogoliubov approximation, it is interesting to
confirm that the dipole modes are in fact unrenormalized in this
approximation.
We therefore construct the exact solutions of the Bogoliubov equations
that correspond to the dipole modes.  These may be useful points of
comparison with numerical calculations of the Bogoliubov spectrum.
Finally, we examine the form of these modes in the Thomas-Fermi
approximation\cite{BP}, which is relevant for large condensates.

\subsection*{Definitions.}
Consider bosons of mass $m$ confined to a three-dimensional, harmonic
potential with spring constants $m\omega_\alpha^2$, where $\alpha =$ $x,$
$y,$ and $z$ labels the three cartesian directions.
The corresponding bare trap frequencies are $\omega_\alpha$.
The (one-body) Hamiltonian of a single particle in the trap is then
\begin{mathletters}
\begin{equation}
{\cal H}_0 \equiv
\sum_{\alpha} \bigg( \frac{p_\alpha^2}{2m} +
\case 1/2 m \omega_\alpha^2 r_\alpha^2 \bigg) ,
\end{equation}
or, equivalently,
\begin{equation}
{\cal H}_0=\sum_{\alpha} \hbar\omega_\alpha (a_\alpha^\dagger a_\alpha +
\case 1/2),
\end{equation}
\end{mathletters}
where we introduce the usual raising and lowering operators
\begin{mathletters}
\label{eq:aadag}\begin{equation}
a_\alpha \equiv
\frac{1}{\sqrt{2}} \bigg(\frac{x_\alpha}{d_\alpha} +
d_\alpha{\partial \over {\partial x_\alpha}}\bigg),
\label{eq:a}
\end{equation}
\begin{equation}
a_\alpha^\dagger \equiv
\frac{1}{\sqrt{2}} \bigg(\frac{x_\alpha}{d_\alpha} -
d_\alpha{\partial \over {\partial x_\alpha}}\bigg),
\label{eq:adag}
\end{equation}
\end{mathletters}
and (as above) $d_\alpha \equiv \sqrt{\hbar/m\omega_\alpha}$.
The operators $a_\alpha$ and $a_\alpha^\dagger$ obey Bose commutation
relations, and they act on one-body states.
They satisfy the familiar relation
\begin{equation}
[{\cal H}_0,a_\alpha^\dagger] = \hbar\omega_\alpha
a_\alpha^\dagger
\label{eq:comm}
\end{equation}
which shows that $a_\alpha^\dagger$ is a raising operator of the
noninteracting system.

Let $V({\bf r} - {\bf r'})$ be the two-body interaction potential between
the particles.  The many-body Hamiltonian is then
\begin{equation}
{\cal H} = \sum_{i=1}^{N} {\cal H}_{0i}
+ \frac{1}{2} \sum_{i\neq j}^N V({\bf r}_i - {\bf r}_j),
\end{equation}
where the index $i$ and $j$ labels individual particles, and $N$ is the
total particle number.  For bosons, we  consider only states that are
symmetric under interchange of particle labels.

\subsection*{Dipole-mode creation operator}
We claim that the (symmetric) operator
\begin{equation}
A_{\alpha}^\dagger \equiv \sum_{i=1}^N a_{\alpha i}^\dagger
\end{equation}
is a raising operator for the many-body Hamiltonian ${\cal H}$.
That is,
\begin{equation}
[{\cal H}, A_\alpha^\dagger] = \hbar\omega_\alpha A_\alpha^\dagger.
\label{big-comm}
\end{equation}
Thus if $|G\rangle$ is the exact many-body ground state of the interacting
system,  then $A_\alpha^\dagger | G \rangle$ is an {\it exact\/} excited
state with excitation energy $\hbar \omega_\alpha$.  Repeatedly
applying $A_\alpha^\dagger$ to any exact eigenstate of ${\cal H}$
creates an equally spaced ladder of exact excited states.
[Note that $(A_\alpha^\dagger)^2$ creates two quanta of the
same elementary excitation and does not yield a distinct new elementary
excitation itself.]

The proof is straightforward.  The commutator Eq. (\ref{big-comm}) is
\begin{equation}
[{\cal H}, A_\alpha^\dagger] =
\sum_{ij} [{\cal H}_{0i},a_{\alpha j}^\dagger]
+\case 1/2 \sum_{ijk} [V_{ij},a_{\alpha k}^\dagger].
\end{equation}
The first set of commutators is easily evaluated using Eq.~(\ref{eq:comm})
and gives
$\hbar\omega_\alpha\sum_i a_{\alpha i}^\dagger$$=$$\hbar\omega_\alpha
A_{\alpha}^\dagger$.  The second set of commutators vanishes identically,
since the terms cancel in pairs.

This operator $A_\alpha^\dagger$ makes good physical sense.  Beginning from
any exact many-body eigenstate, we raise each particle in turn by a single
quantum, and then superimpose the resulting states to produce a symmetric
wave function.  The proof is similar to the demonstration that the cyclotron
frequency of an interacting, translationally invariant system is
unrenormalized, {\it i.e.,} Kohn's theorem\cite{Kohn}.

\subsection*{Dipole mode in the Bogoliubov approximation.}
Is the same dipole mode also present as an exact solution of the
Bogoliubov equations for harmonically trapped bosons?  The answer
to this question is not completely obvious, since the Bogoliubov
approximation assumes the existence of a condensate, which provides
a provides a preferred reference frame (usually, but not necessarily,
at rest).  We will show that the spectrum of the dilute Bose gas in
the Bogoliubov approximation possesses {\it exact\/} excited states
that are simply related to the dipole modes discussed above.

Let $\Psi({\bf r})$ be the condensate wave function of the interacting
system, which is an exact solution of the Gross-Pitaevskii equation
[Eq.~(\ref{eq:GP})]:
\begin{equation}
({\cal H}_0 + g|\Psi|^2) \Psi = \mu \Psi,
\label{eq:gpagain}
\end{equation}
where $ g = 4\pi a \hbar^2/m$ characterizes the strength of the
interparticle potential and $\mu$ is the chemical potential.
We claim that the excited state
\begin{equation}
\pmatrix{u_\alpha({\bf r})\cr v_\alpha({\bf r})} =
\pmatrix{a_\alpha^\dagger \Psi({\bf r})\cr a_\alpha
\Psi^*({\bf r})}
\label{eq:uv}
\end{equation}
is then an exact solution of the Bogoliubov equations with excitation
frequency equal to the bare trap frequency $\omega_\alpha$.
There is one such mode for each coordinate direction.

Note that for the noninteracting case ($g=0$), $a_\alpha$ annihilates the
condensate wave function $\Psi$, which is then simply the ground state
$\Psi_0$ of the harmonic potential.  Equation (\ref{eq:uv}) then reduces to
\begin{equation}
\pmatrix{u_\alpha({\bf r})\cr v_\alpha({\bf r})} =
\pmatrix{a_\alpha^\dagger\Psi_0({\bf r})\cr 0}
\qquad \text{(noninteracting)}.
\end{equation}
As expected, exact quasiparticle states are created by adding a
particle to the first excited state of the trap.

It is straightforward to prove that Eq. (\ref{eq:uv}) is an exact
solution of the Bogoliubov equations for {\it any} coupling strength $g$.
The proof proceeds  by the explicit demonstration
that the $u_\alpha$, $v_\alpha$ of  Eq.~(\ref{eq:uv}) satisfy the
coupled Bogoliubov equations~(\ref{Bog}).
We use the elementary properties of $a_\alpha$ and $a_\alpha^\dagger$,
and the fact that $\Psi$ satisfies the Gross-Pitaevskii equation
[Eq. (\ref{eq:gpagain})].
The argument explicitly relies on the harmonic character of the trap
$V({\bf r})$, with $a_\alpha^\dagger$ and $a_\alpha$ as raising and
lowering operators for the single-particle problem; it cannot
be extended to problems without a simple ladder of noninteracting levels.

We now confirm that the first of the two coupled Bogoliubov
equations Eq.~(\ref{eq:Boga}) is satisfied by our proposed solution
Eq.~(\ref{eq:uv}); the second equation Eq.~(\ref{eq:Bogb}) is simply
the complex conjugate of the first.  Substituting Eq.~(\ref{eq:uv})
into Eq.~(\ref{eq:Boga}), we need to show that
\begin{equation}
({\cal H}_0 + g|\Psi|^2 - \mu)\, a_{\alpha}^{\dagger} \Psi +
g|\Psi|^2 a_\alpha^\dagger \Psi - g \Psi^2 a_\alpha \Psi^*
\stackrel{?}{=} \hbar\omega_\alpha a_\alpha^\dagger \Psi.
\label{eq:upper1}
\end{equation}
Multiplying the Gross-Pitaevskii equation~(\ref{eq:GP}) by
$a_\alpha^\dagger$ yields
\begin{equation}
a_\alpha^\dagger ({\cal H}_0 + g|\Psi|^2 - \mu) \Psi = 0.
\label{eq:upper2}
\end{equation}
Subtracting  Eq.~(\ref{eq:upper2}) from Eq.~(\ref{eq:upper1}) then
reduces our problem to showing that
\begin{equation}
[{\cal H}_0, a_\alpha^\dagger] \Psi +
g[\,|\Psi|^2, a_\alpha^\dagger] \Psi +
g|\Psi|^2 a_\alpha^\dagger \Psi - g \Psi^2 a_\alpha \Psi^*
\stackrel{?}{=} \hbar\omega_\alpha a_\alpha^\dagger \Psi.\label{eq:upper3}
\end{equation}
From Eq.~(\ref{eq:comm}), the first commutator is simply
$[{\cal H}_0,a_\alpha^\dagger] =\hbar\omega_\alpha a_\alpha^\dagger$,
which cancels the right hand side of Eq.~(\ref{eq:upper3}).

Finally, our task is reduced to showing that
\begin{equation}
[\,|\Psi|^2, a_\alpha^\dagger] \,\Psi +
|\Psi|^2 a_\alpha^\dagger \Psi -  \Psi^2 a_\alpha \Psi^*
\stackrel{?}{=} 0;
\end{equation}
this result is easy to confirm with the explicit form of $a_\alpha^\dagger$
and $a_\alpha$ in Eqs.~(\ref{eq:a}) and (\ref{eq:adag}).

Note that this construction is entirely general.  It makes no assumption
that $\Psi$ is real, and thus holds for {\it any\/} solution of the
Gross-Pitaevskii equation (\ref{eq:GP}), including those describing
vortices\cite{DBEC}.
For any self-consistent condensate, the lowest dipole modes will have the
bare oscillation frequencies.

\subsection*{Nature of the dipole mode.}
Excitation of the dipole mode leads to an oscillatory density-fluctuation
amplitude that is one of the normal modes described in Eq.~(\ref{linab}).
For simplicity, we consider explicitly only the case of a condensate with
real $\Psi$, {\it i.e.,} a stationary condensate.

Equations (\ref{linab}) then become
\begin{eqnarray}
\delta \rho_\alpha  &=& \Psi\,(u_\alpha - v_\alpha )
=\Psi\big(a_\alpha^\dagger\Psi - a_\alpha\Psi\big) \cr\noalign{\smallskip}
&=& \sqrt 2\Psi d_\alpha \frac{\partial \Psi}{\partial r_\alpha}
\propto \frac{\partial n_0}  {\partial r_\alpha}.
\label{eq:dens-fluc}
\end{eqnarray}
This expression confirms that the quasiparticle mode created by the linear
combination of $a_\alpha^\dagger$ and $a_\alpha$ corresponds to a uniform
displacement of the condensate in the $\alpha$th direction.
Similarly, the velocity potential associated with this condensate motion is
[see Eq.~(\ref{linab})]
\begin{eqnarray}
\delta\Phi_\alpha &=& \frac{\hbar}{2mi\Psi}\,(u_\alpha+v_\alpha)
= \frac{\hbar}{2mi\Psi}\,(a_\alpha^\dagger\Psi
+a_\alpha\Psi)\cr\noalign{\smallskip}
&=& \frac{-i\hbar}{\sqrt2 m} \, \frac{r_\alpha}{d_\alpha};
\end{eqnarray}
the corresponding velocity ${\bf v} = \nabla \delta\Phi_\alpha$
lies along the displacement with  constant amplitude and is
$\frac{1}{2}\pi$ out of phase relative to the density fluctuation
Eq. (\ref{eq:dens-fluc}) due to the factor of $i$.

\subsection*{Thomas-Fermi approximation}
Finally, we examine the dipole mode in the Thomas-Fermi limit, where the
Gross-Pitaevskii equation can be solved exactly.
For simplicity we consider an isotropic trap, although the same results
hold for anisotropic traps of the sort used in Refs. \cite{SR,Edw}.
The condensate wave function is then given by
\begin{equation}
\Psi_{TF}({\bf r}) \propto\sqrt{R_0^2 - r^2},
\end{equation}
where  $R_0 \propto N_0^{1/5}$ is the size of the condensate.
Using Eq.~(\ref{linv}) and the  operators from
Eqs.~(\ref{eq:aadag}), we find
\begin{mathletters}
\label{eq:TFuv}
\begin{equation}
u_\alpha({\bf r}) =
a_\alpha^\dagger \Psi({\bf r}) \propto r_\alpha \bigg[\sqrt{R_0^2-r^2} + {1
\over \sqrt{R_0^2 - r^2}}
\bigg]
\end{equation}
\begin{equation}
v_\alpha({\bf r}) =
a_\alpha \Psi({\bf r}) \propto r_\alpha\bigg[\sqrt{R_0^2-r^2} - {1 \over
\sqrt{R_0^2 - r^2}} \bigg]
\end{equation}
\end{mathletters}
Although $\delta\rho$ and $\delta\Phi$ remain finite everywhere, both both
$u_\alpha$
and $v_\alpha$ diverge at the perimeter of the condensate.
This pathology evidently arises from the extreme Thomas-Fermi
limit, for it reflects the singular behavior of $\Psi_{TF}$ at the
condensate boundary.  Any finite interparticle interaction strength
$g$ renders the condensate wave function $\Psi$ differentiable everywhere,
and the exact $\Psi$ vanishes smoothly for $r \rightarrow \infty$ \cite{DPS};
then $u_\alpha$ and $v_\alpha$ also vanish smoothly.
Despite the separate divergences, the approximate two-component Bogoliubov
state in the TF limit in Eq.~(\ref{eq:TFuv}) remains normalizable,
since we require only that
$\int_0^\infty r^2\,dr\,\big(|u_\alpha|^2- |v_\alpha|^2\big) = 1$.

\end{document}